\documentclass{aastex62}

\received{May 25, 2019}
\submitjournal{ApJL}

\shorttitle{Acoustic cut-off frequency}
\shortauthors{Jefferies et al.}

\begin{document}

\title{Observed local dispersion relations for magneto-acoustic-gravity waves in the Sun's atmosphere: Mapping the acoustic cut-off frequency}

\correspondingauthor{Stuart M. Jefferies}
\email{sjefferies@gsu.edu}

\author{Stuart M. Jefferies}
\affiliation{Georgia State University, 25 Park Place, Atlanta, GA 30303, USA}
\affiliation{University of Hawaii, 34, Ohia Ku Street, Pukalani, HI 96768, USA}
\nocollaboration

\author{Bernhard Fleck}
\affil{ESA Directorate of Science, Operations Department, c/o NASA/GSFC Code 671, Greenbelt, MD 20071, USA}

\nocollaboration
\author{Neil Murphy}
\affiliation{Jet Propulsion Laboratory, Pasadena, CA 91109, USA}

\nocollaboration
\author{Francesco Berrilli}
\affiliation{Universit\`{a} di Roma Tor Vergata, Via Ricerca Scientifica 1, I-00133, Roma, Italy}



\begin{abstract}
 We present the observed local dispersion relations for magneto-acoustic-gravity waves in the Sun's atmosphere for different levels of magnetic field strength. We model these data with a theoretical local dispersion relation to produce spatial maps of the acoustic cut-off frequency in the Sun's photosphere. These maps have implications for the mechanical heating of the Sun's upper atmosphere, by magneto-acoustic-gravity waves, at different phases of the solar magnetic activity cycle.

\end{abstract}



\section{Introduction} 
\label{sec:intro}

The mechanisms behind the radiative losses of the atmospheres of stars with outer convection zones, such as the Sun, are still not fully understood \citep{2015RSPTA.37340269D}. One mechanism under consideration is mechanical heating by acoustic waves generated by the turbulent convection near the star's surface. When these waves propagate into the stars atmosphere, they can deposit the convective energy they are carrying via shock formation. However, whether or not acoustic waves can propagate in a stratified medium such as a stable atmosphere is determined by the acoustic cut-off frequency of the atmosphere \citep{1909PLMS...7..122L}. Waves with frequencies below the cut-off frequency are evanescent (i.e., non-propagating) while waves with frequencies above the cut-off can freely propagate. 

Theoretically, the acoustic cut-off frequency approximately scales as $g \sqrt{\mu/T_{eff}}$ where $T_{eff}$ is the effective temperature, $g$ is the gravity, and $\mu$ is the mean molecular weight, with all values measured at the surface. Hence direct measurement of the cut-off frequency can, in principle, be used to constrain the fundamental stellar parameters. Having said that, the acoustic cut-off frequencies for stars (e.g., roAp stars \citep{1998A&A...335..954A}), are currently theoretically determined from stellar models. However, for the Sun we are able to measure the cut-off frequency and we have a mechanism to check our theoretical expectations.

Although the cut-off frequency is expected to be a locally defined quantity \citep{2018ApJ...869...36C}, only its variation with height has received any attention \citep{2018A&A...617A..39F, 2016ApJ...827...37M, 2016ApJ...819L..23W}. To the best of our knowledge, no one has examined how the cut-off frequency varies spatially in latitude and longitude.

\vspace{0.2in}

\section{Observations and results}
\label{obs_res}

For the analysis presented here we used 11 hours of line-of-sight Doppler velocity data acquired on 21 January 2017 from 07:01:30 UT to 18:01:30 UT using the MOTH II (hereafter referred to as MOTH) \citep{2018IAUS..335..335F} and SDO/HMI \citep{2012SoPh..275..207S, 2012SoPh..275..229S} instruments. In addition, to probe the lower photosphere, we also used 11 hours of Doppler velocities derived from the first and second Fourier coefficients calculated from the individual HMI filtergrams \citep{2012SoPh..278..217C,  2014SoPh..289.3457N} acquired on 24 August 2010 from 00:00:00 UT to 11:00:00 UT (i.e. during solar minimum conditions). The Doppler signal derived from the phase of the second Fourier coefficients is closer to a line core Doppler signal than the standard HMI Doppler signal, which is derived from the phase of the first Fourier coefficients. Applying the procedure of \citet{2011SoPh..271...27F} to the Doppler signal derived from the second Fourier coefficients suggests a formation height that is about 40\,km higher than that of the standard HMI Doppler velocities, i.e. at around 180\,km vs 140\,km (taking into account the limited spatial resolution, which shifts the apparent line formation by about 40 to 50\,km to larger heights). 

The MOTH II instruments use the Na 589 nm and K 770 nm solar absorption lines to sample the upper photosphere/lower chromosphere region of the Sun's atmosphere, while the HMI instrument uses the Fe 617 nm line to sample the lower photosphere. The MOTH II observations were acquired at a cadence of 5 seconds but were integrated to a 45 second cadence to be commensurate with the HMI sampling rate.  Both the MOTH II and the HMI data were spatially binned to yield an effective pixel size of 1.27 arcsec.

\begin{figure}[!ht]
\centering
\plottwo{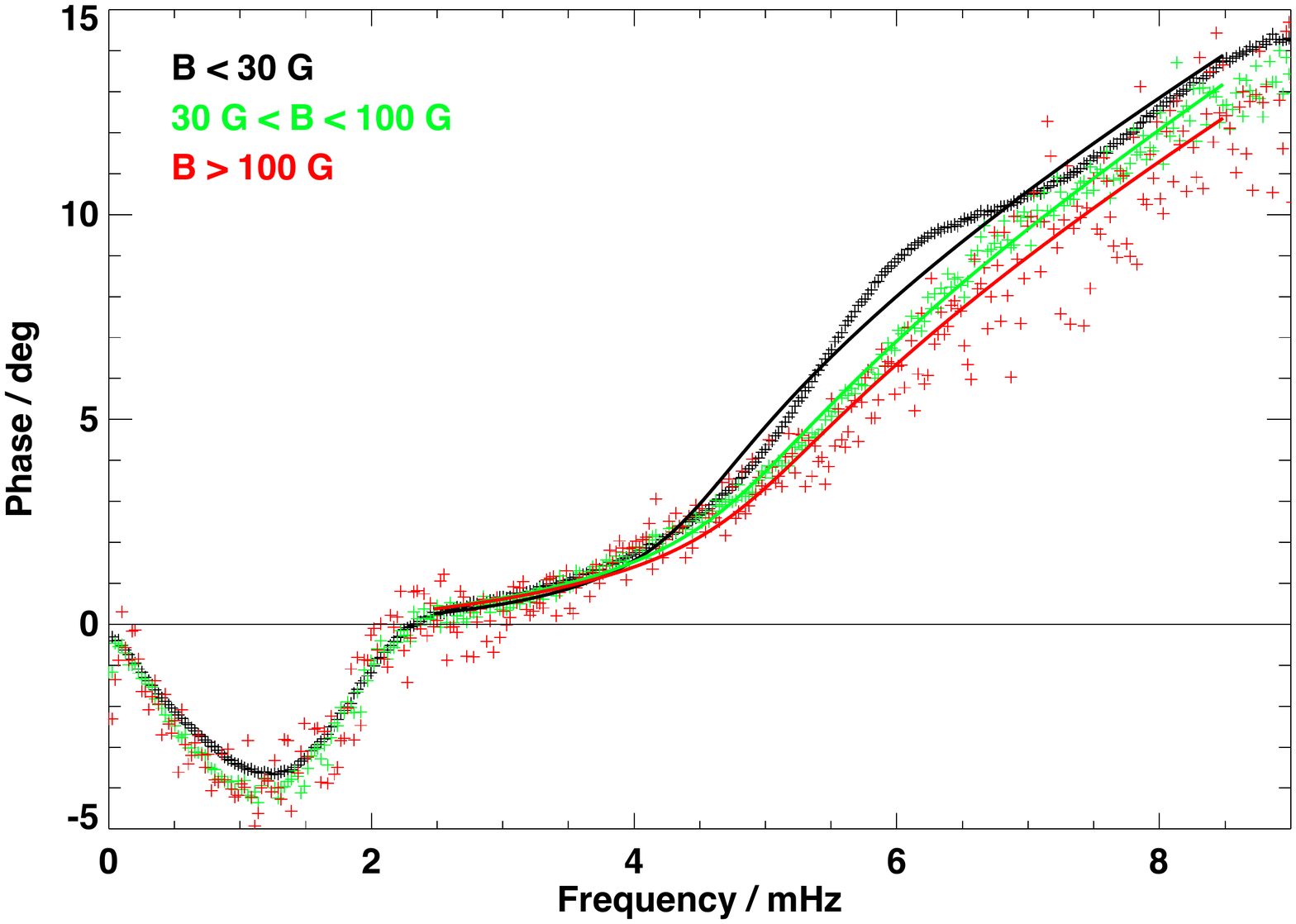}{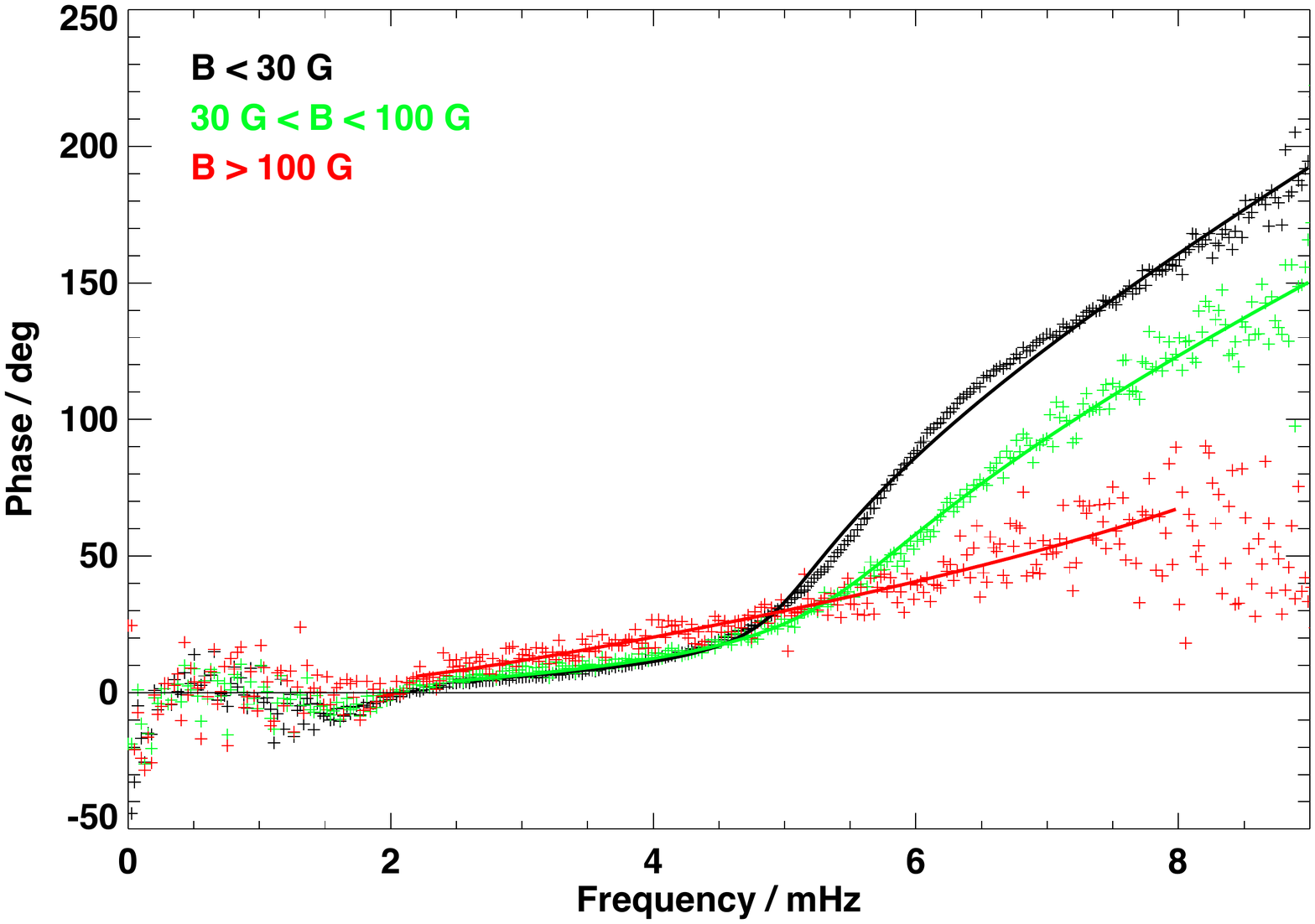}
\caption{The observed phase differences (denoted by crosses) in three different regions of magnetic field strength, $\mid B \mid \le$ 30~G (black), 30~G $< \mid B \mid \le$ 100~G (green), and $\mid B \mid >$ 100 G (red), and at two different heights in the Sun's photosphere. In this Letter we consider the phase differences above 2 mHz. Fits using Equations (\ref{eqn:1} and \ref{eqn:kz1}) are overlaid (thick solid lines). 
The phase difference spectra for the 2010 Fe - Fe observations (lower photosphere) are shown on the left and represent the mean of 977,253, 64,599, and 6,764 individual phase difference spectra for the black, green and red data, respectively. 
The phase difference spectra for the 2017 Na - Fe observations (upper photosphere) are shown on the right and represent the mean of 616630, 17298 and 6072 individual phase difference spectra for the black, green and red data, respectively. 
}
\label{fig:Dispersion}
\end{figure}

\par

We computed the crosspower spectra $CP(\omega) = F_{V_1}(\omega) \cdot F_{V_2}(\omega)^\star$ with $F$ denoting the Fourier transform of the measured velocity signals $V_1$ and $V_2$ between the 2017 MOTH and HMI data sets (i.e., Na - Fe, K-Fe, Na-K) and between the two 2010 HMI data sets (i.e., Fe - Fe) for each spatial pixel in a  1060 arcsec x 1060 arcsec region near disk center. For the 2010 HMI data sets we used a smaller region of 518 arcsec x 518 arcsec (center 1024 x 1024 pixels of the full 4k x 4k resolution HMI Dopplergrams). 
We then averaged the crosspower spectra over different ranges of magnetic field strength and calculated the phase difference spectra  $\Delta \phi(\omega)=\arctan(Imaginary(CP(\omega))/Real(CP(\omega))$ to provide the curves shown in Fig. \ref{fig:Dispersion}. 

To model the observed phase difference we start by defining the vertical component of the velocity induced by a upward propagating plane wave of angular frequency $\omega$ and angular wave number $k$, where $k^2 = k_x^2 + k_z^2$, $k_x$ and $k_z$ are the horizontal and vertical components, respectively, as the real part of
\begin{equation}
    v_z(x,z,t) = U_z(0) \exp{\kappa z} \exp{\left[i(\omega t - k_z z - k_x x)\right]}
    \label{rd2}
\end{equation}
where $U_z(0)$ is the amplitude of the wave at height $z=0$ and $\kappa$ is related to the scale height of the atmosphere (see below).
The phase difference $\Delta\phi$ is given by

\begin{equation}\Delta \phi = k_z  \Delta z = \omega \Delta t_{ph},
\label{eqn:1}
\end{equation}
where $\Delta z$ is the height difference $(z_2 - z_1)$ between the observations and $\Delta t_{ph}$ is the phase travel time of the wave between the observations. Equation (\ref{eqn:1}) shows that the observed variation of $\Delta\phi$ with frequency provides a measure of the local magneto-acoustic-gravity wave dispersion relation for upward propagating waves.

We define the magnetically quiet Sun as being those spatial locations where the unsigned line-of-sight magnetic flux from HMI 
$\mid B\mid < 30$ G, and model the corresponding measured phase difference spectra, above 2 mHz, using Equation (\ref{eqn:1}) and $k_z$ as defined by the dispersion relation for acoustic-gravity waves in an isothermal stratified atmosphere with constant radiative damping \citep{1972A&A....17..458S}, i.e., 
\begin{equation}
k_z = \pm \left[\frac{a}{2} + \frac{\sqrt{a^2 + b^2}}{2} \right]^{\frac{1}{2}}
\label{eqn:kz1}
\end{equation}

\begin{equation}
\kappa =\left(\frac{1}{2H}\right) - \left[\frac{-a}{2} + \frac{\sqrt{a^2 + b^2}}{2} \right]^{\frac{1}{2}}
\label{eqn:kappa}
\end{equation}

where
\begin{equation}
    a = {{\omega^2 - \omega_{ac}^2}\over{c_s^2}} + k_x^{2} \left( {{N^2}\over{\omega^2}} - 1\right)
    - {{1}\over{1+\omega^2\tau_R^2}}
    {{N^2}\over{\omega^2}}
    \left(k_x^2 - {{\omega^4}\over{g^2}} \right)
\end{equation}
and
\begin{equation}
    b =  {{\omega \tau_R}\over{1+\omega^2\tau_R^2}}
    \left({{N^2}\over{\omega^2}}\right)
    \left(k_x^2 - {{\omega^4}\over{g^2}} \right).
\end{equation}
Here $\omega_{ac}$ is the acoustic cut-off frequency, $N$ is the buoyancy frequency, $\tau_R$ is the radiative damping time, $c_s$ is the local sound speed, $\gamma$ is the adiabatic exponent, and $g$ is the acceleration due to gravity.

The variables in the model are $\left(c_s, \Delta z, \omega_{ac}, \tau_R \right)$, and the buoyancy frequency is obtained through the identity
\begin{equation}
 N = (\gamma - 1)^{\frac{1}{2}}{{g}/{c_s}}. 
 \end{equation}

In the photosphere $g=274 ~\rm{m/s^2}$ and $\gamma$ is taken to be $5/3$. We set $k_x = 0$ so as to only consider vertically propagating waves.

\par
\begin{figure}[!ht]
\centering
\includegraphics[width=1.0\textwidth]{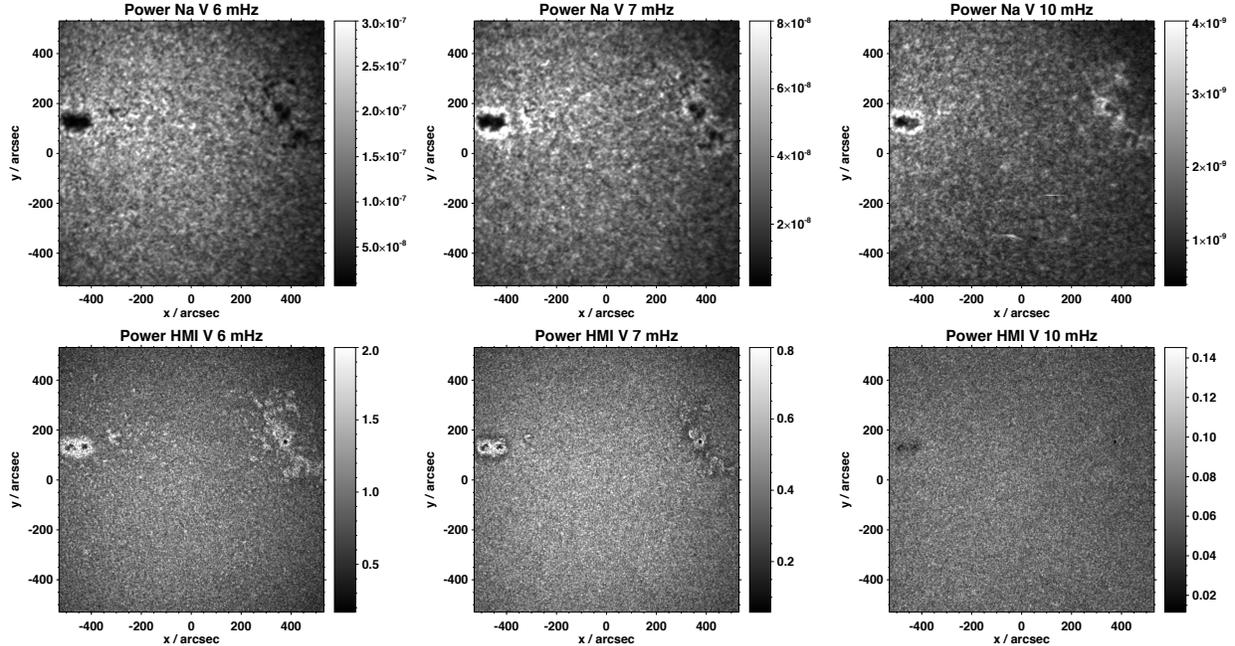} 
\caption{The observed power maps at select frequencies above 5 mHz for the 2017 HMI (bottom row) and MOTH Na observations (top row). The maps at 6 mHz show strong power halos around the magnetic regions in the HMI map but not in the MOTH Na map. The maps at 7 mHz show strong power halos in both the HMI and MOTH Na maps. In addition the HMI map shows regions of suppressed power just outside of the power halos, that is spatially commensurate with the power enhancements seen in the MOTH Na map. Both of these observations are seen in the modeling of \citep{2015ApJ...801...27R} and point to the existence of downward propagating waves (see also, e.g., \citet{2007A&A...471..961M} and \citet{2009A&A...506L...5K}).}
\label{fig:power2017}
\end{figure}

 The results are shown in Fig. \ref{fig:Dispersion}. As can be seen in both panels of this figure the theoretical phase difference curve generated using Equations (\ref{eqn:1} and \ref{eqn:kz1}) models the overall behavior of the observed phase difference data well. However it does not display sufficient flexibility to model the small "wave" in the observed phase difference spectra that is centered just above 6 mHz. (We note, we also see the same behavior in a MOTH K - HMI Fe phase difference plot which is not shown here because of space limitations.) There is clearly some physics that is not being addressed in Equation (\ref{eqn:1}). To address this we turn to the observed spatial maps of power at high frequencies ($>$ 5 mHz), for the MOTH II Na and HMI data sets (see Fig. \ref{fig:power2017}). 

The behavior in these maps has been well described by the recent modeling of \citet{2015ApJ...801...27R} who showed that the "halos" of increased power near active regions is caused by waves reflected from the region where the Alfv\'{e}n speed, $a$, is commensurate with the sound speed, $c_s$. Basically, fast acoustic waves with frequencies above the acoustic cut-off frequency propagate up to the $a=c_s$ layer where they are transmitted and converted into slow acoustic waves and fast magnetic waves, respectively. The slow acoustic waves continue their journey upward, guided by the magnetic field lines. The fast magnetic waves, however, are refracted and then finally reflected at the height where $\omega/k_x = a$ is met \citep{2015ApJ...801...27R}. These reflected waves then transmit and convert into slow magnetic and fast acoustic waves, respectively.  The former are field guided, similar to the slow acoustic waves in the region $a >> c_s$. \citet{2015ApJ...801...27R} showed that when wave reflection is suppressed in their model, the power halos disappear. This provides strong evidence that there are downward propagating waves in the solar atmosphere at frequencies above the acoustic cut-off frequency.
 We note that additional support for the presence of a reflecting layer in the atmosphere is provided in the modulation of the time-distance diagrams for waves with frequencies $>$ 5 mHz \citep{1997ApJ...485L..49J}.
 
 With this in mind, we find that if we model the velocity signal as consisting of upward and downward waves with $k_x= 0)$, i.e.,
 \begin{equation}
    v_z(z,t) = U_z(0) \exp{(\kappa z)} \left( \exp{\left[i(\omega t - k_z z)\right]} + \alpha \exp\left[i\psi(\omega)\right]
    \exp{\left[i(\omega t + k_z z)\right]} \right)
    \label{rd}
\end{equation}
 where $\alpha \le 1$ is related to the coefficient for the fast acoustic to fast magnetic (and vice versa) wave conversion process, and $\psi(\omega)$ represents a frequency dependent phase change of the wave on reflection (e.g., see Equation (2) of \citet{1991MNRAS.251..427W}), then we can produce a "wave" in the observed phase difference spectra that is similar to one we observe.  This suggests that accurate modeling of the phase difference spectra should therefore allow for both upward and downward travelling waves. 
 
Now Equation (\ref{eqn:kz1}) is the dispersion relation for acoustic gravity waves in an isothermal atmosphere. If we look at the dispersion relations for waves in a magnetic atmosphere, such as equation (12) in \citet{2006MNRAS.372..551S}, we expect 
two magneto-acoustic waves that are acoustic-like in their behavior (see equations (14) and (16) in \citet{2006MNRAS.372..551S}). One is known as the "slow" acoustic wave and it exists above the $a = c_s$ layer, and the latter is known as the "fast" acoustic wave and it exists below the $a = c_s$ layer. (The "fast" and "slow" waves above and below the $a=c_s$ layer, respectively, are magnetic in character.)  For $k_x=0$, and the two limiting cases $a >> c_s$ and $c_s>> a$ we have

\begin{equation}
    \omega^2 - \omega_{0}^2 = c_{ph}^2 k_z^2
    \label{eqn:Schunker}
\end{equation}
 where $\omega_0= \omega_{ac} \eta$ and $c_{ph} = c_s \eta$ are the modified acoustic cut-off frequency and phase speed, respectively, with $\eta = \cos\theta$ for $a>>c_s$ and $\eta = 1$ for $c_s >> a$. Here $\theta$ is the inclination angle of the magnetic field to the vertical. In general, $\eta$ depends on $(c_s, a, \theta)$ in a complicated way  \citep{1977A&A....55..239B}. 

If we neglect the effects of radiative damping, and only consider vertically propagating waves (i.e. $k_x=0$), then we can see that Equations (\ref{eqn:kz1}) and (\ref{eqn:Schunker}) have the same form for the dispersion relation. To test if this general model for the dispersion relation can be applied to regions where $a \sim c_s$  we explored the use of Equations (\ref{eqn:1} and \ref{eqn:kz1}) to model the phase difference curves for regions where $\mid B \mid$ is larger than our "quiet Sun" value of 30 G. In this case, however, the parameters associated with $c_s$ and $\omega_{ac}$ represent $c_{ph}$ and $\omega_0$. The results, shown by the green and red curves in Figure \ref{fig:Dispersion}, clearly demonstrate the acoustic-gravity dispersion model works astonishingly well, even for the highest magnetic field strengths.
 
\begin{figure}[!ht]
\centering


\begin{minipage}{0.45\textwidth}
\centering
\includegraphics[width=1\textwidth]{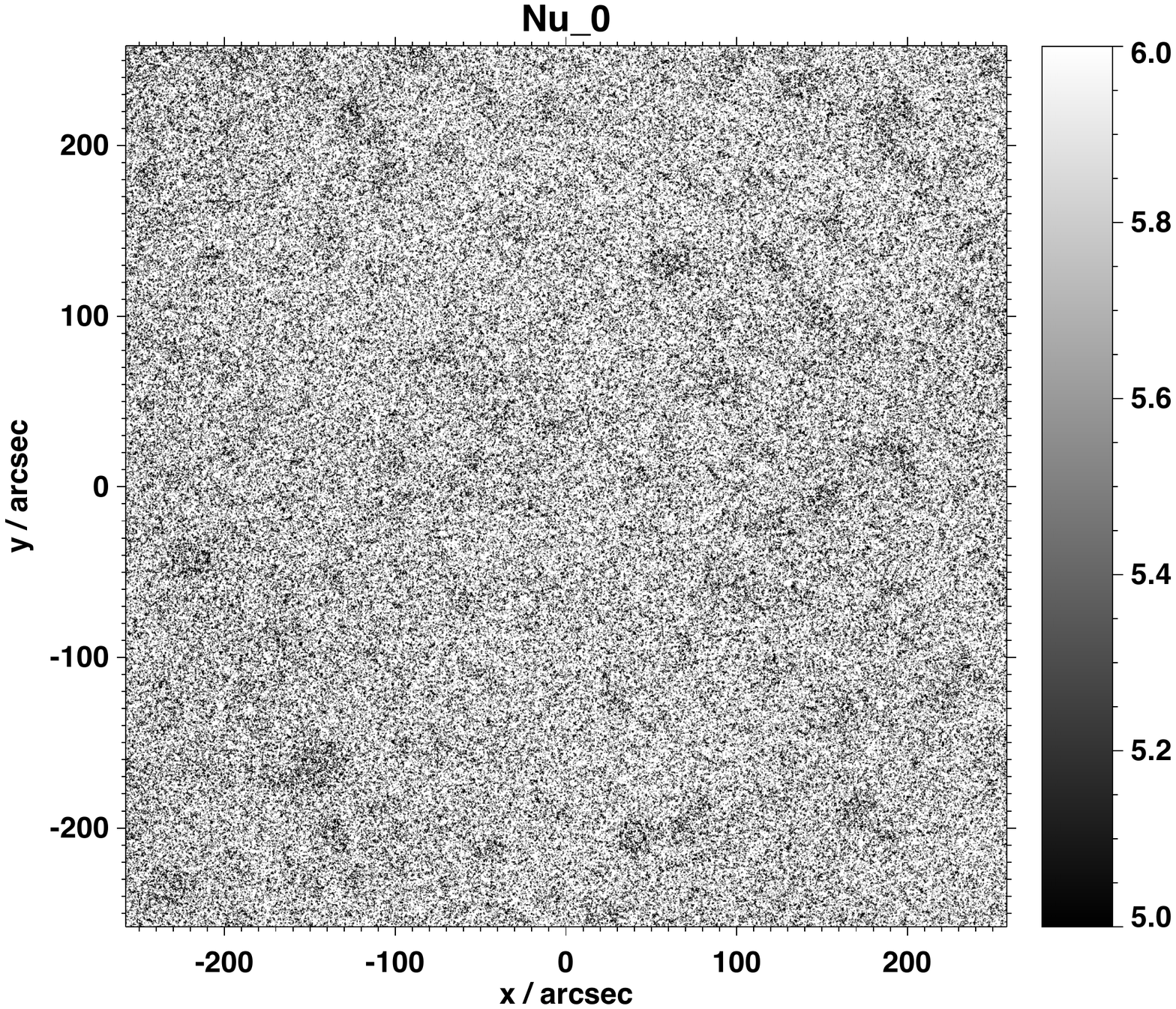}
\end{minipage}\hfill
\begin{minipage}{0.45\textwidth}
\centering
\includegraphics[width=1\textwidth]{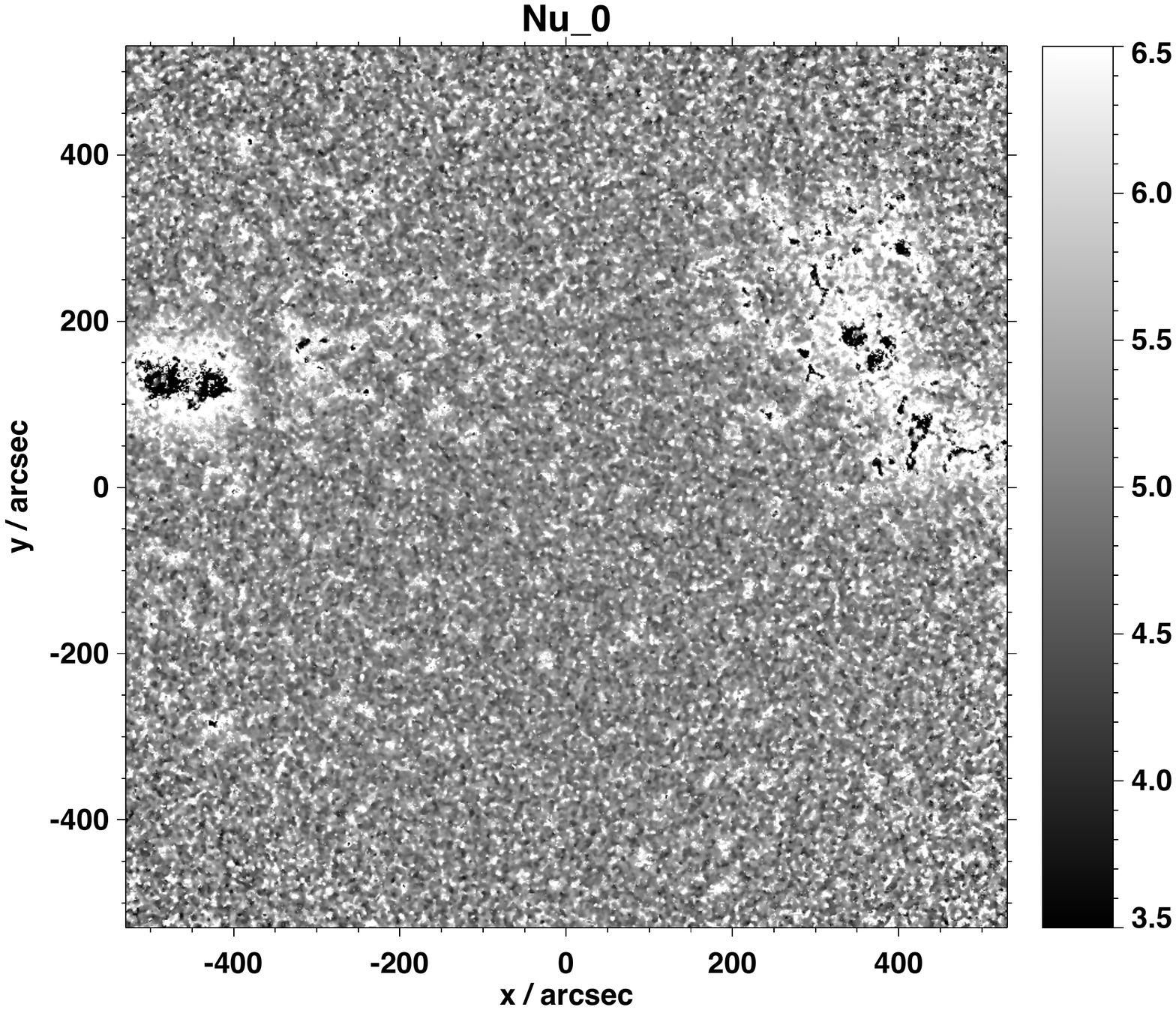}
\end{minipage}

\begin{minipage}{0.45\textwidth}
\centering
\includegraphics[width=1\textwidth]{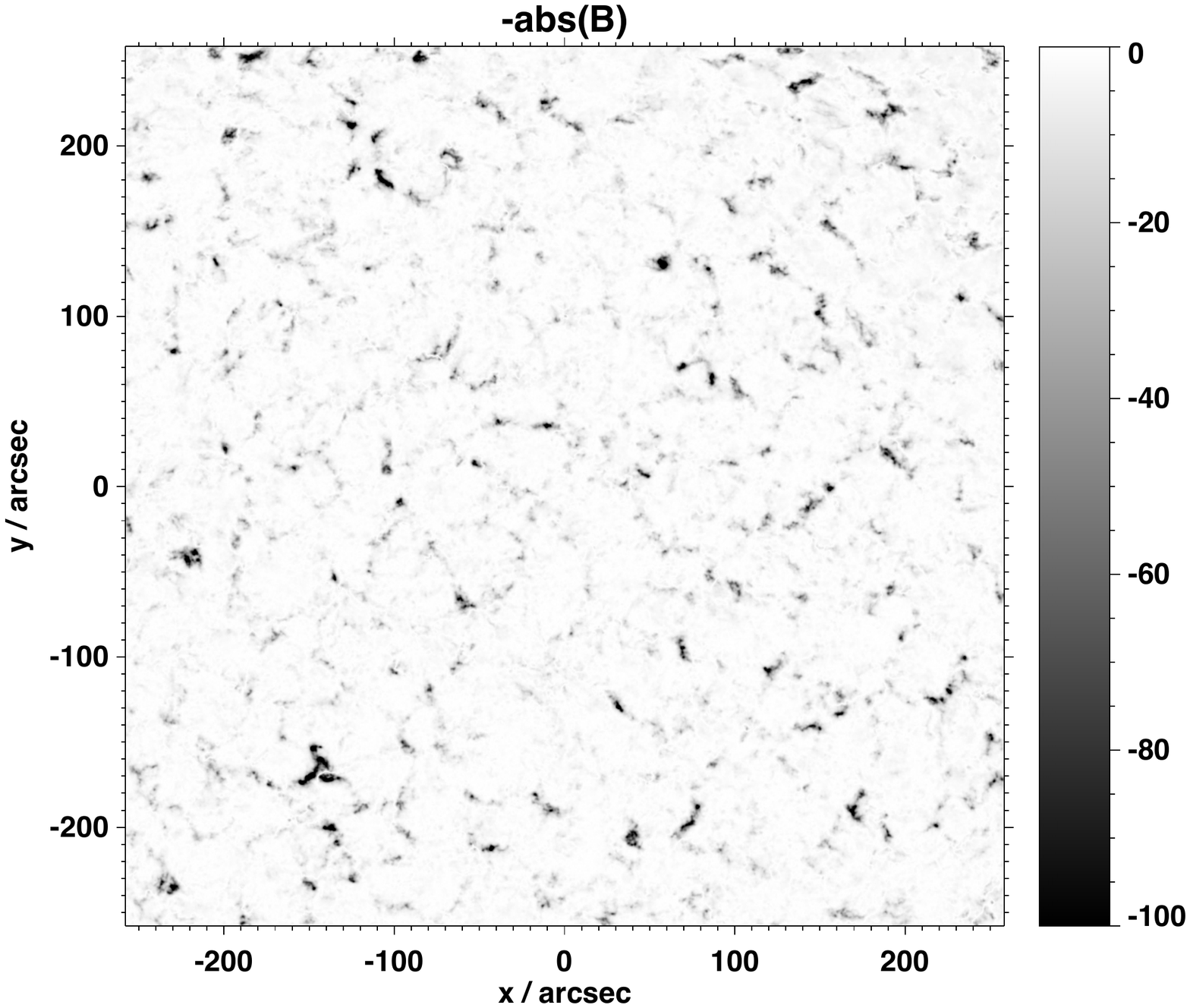}
\end{minipage}\hfill
\begin{minipage}{0.45\textwidth}
\centering
\includegraphics[width=1\textwidth]{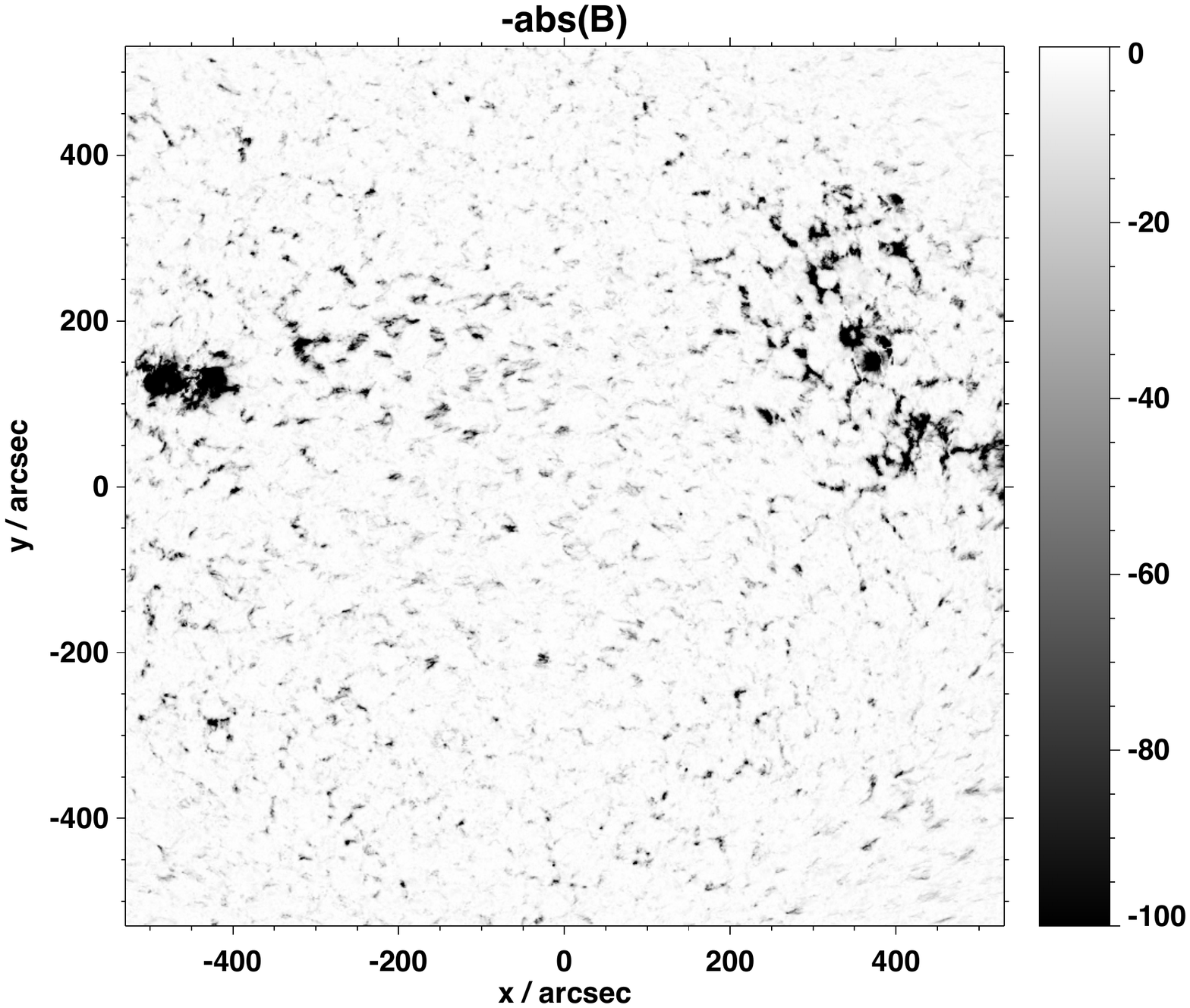}
\end{minipage}


\caption{Top row: The modified acoustic cut-off frequency, $\nu_0 = \nu_{ac} \cos\theta = \omega_{ac}/2\pi \cos\theta$ at different heights in the solar atmosphere, in mHz. Left: 2010 Fe-Fe data, i.e. lower photosphere; mean height approximately 160 km. Right: 2017 Na-Fe data, i.e. upper photosphere; mean height approximately 350 km.
Bottom row: The average unsigned magnetic field at the height of the HMI observations over the 11-hour observational periods, in G. Left: For the 2010 observations. Right: For the 2017 observations. 
As expected,  the internetwork magnetic field in both magnetograms shows spatial frequencies commensurate with supergranulation scales (the field-of-view for the Na-Fe data is approximately twice that of the Fe-Fe data). Careful examination of the Fe-Fe cut-off frequency map, which represents the quiet magnetic sun, shows decreased $\nu_{0}$ in and around the locations of larger magnetic field. This local reduction in $\nu_{0}$ is consistent with the presence of magneto-acoustic portals \citep{2006ApJ...648L.151J, 2019ApJ...871..155R}. These portals allow waves that are normally trapped, to escape and propagate in the atmosphere where they can deposit their energy and contribute to the local heating of the atmosphere.} 
\label{fig:HMI-cut-off}
\end{figure}

Since the discrepancies between our theoretical model and the observed phase difference curves are only apparent for the lowest values of magnetic field strength, and even then they are not dramatic, and the individual phase difference spectra for each pixel in our observations are considerably noisier than in the mean spectra, we used Equations (\ref{eqn:1} and \ref{eqn:kz1}) to provide spatial maps of $\omega_0$, $c_{ph},~\Delta z,$ and $\tau_R$. Because the cross spectra are noisy, we smooth them using a Gaussian of full width at half-maximum of 0.5 mHz before we compute the phase difference signal. 
Although our analysis provides spatial maps of each of the fitted parameters, in this Letter we focus on the variation of $\omega_0$ in the lower solar atmosphere (see Figure \ref{fig:HMI-cut-off}). We remind the reader that $\omega_0$ represents a lower limit of $\omega_{ac}$.


We note that while there have been a number of other studies looking at the acoustic cut-off frequency, they have been
focused on its variation with height in the atmosphere. Here we are concerned with the spatial variation. We should emphasize, though, that we are not referring to the variation at a fixed geometrical height. The solar atmosphere is highly corrugated and the line contribution functions vary considerably (by several hundred kilometers) over the solar surface (cf., e.g., Figs. 2 and 3 in \citet{2011SoPh..271...27F}), in particular also in and around magnetic structures. What we display in Fig. \ref{fig:HMI-cut-off} are maps of the modified acoustic cut-off frequency between the formation heights of the Doppler signals derived from the Fe and Na lines. Their spatial variations may therefore be a mix of horizontal and vertical variations of the measured quantities. Any study using absorption lines is affected by this effect, and there is no technique that would allow mapping derived physical quantities to a certain, fixed geometrical height.

The maps in Fig. \ref{fig:HMI-cut-off} show values for $\nu_{0}= \omega_0/2\pi$ in magnetic regions that are less than that found in regions of magnetically quiet Sun (defined here as regions where $\mid B \mid  < \rm{30~G}$), surrounded by values of $\nu_{0}$ that are significantly larger than the quiet Sun value. The former has been observed before \citep{2018A&A...617A..39F}, however, to the best of our knowledge, the latter hasn't. In retrospect, the presence of values of $\nu_{0}$ larger than the Quiet Sun value was apparent in Fig. 3 of \citet{2006ApJ...648L.151J} where the phase travel-time curve for the low-$\beta$ environment (where $\beta$ is the ratio of gas pressure to magnetic pressure) and small inclination angle is shifted to higher frequencies than the phase travel-time curve for the quiet, ``non-magnetic'' Sun.

\section{Discussion}

The finding of regions in the Sun's atmosphere where the acoustic cut-off frequency is larger than its value in the quiet Sun may provide a connection with other seemingly disparate observations of the Sun's behavior. 
An increase in the value of $\nu_{ac}$ above its value in the quiet Sun would extend the range of frequencies over which waves are trapped beneath the photosphere: it would also cause the effective level of wave reflection for the lower-frequency waves to move inwards. That is, it would result in a shortening of the acoustic cavity. This is because the repulsive properties of the acoustic barrier get stronger \citep{2004ApJ...613L.185F}. As \citet{2004ApJ...613L.185F} pointed out, there is some evidence from global helioseismic measurements for such a picture: inversions of the even-splitting rotation coefficients measured from data obtained at different activity levels show the effective acoustic radius of the Sun to be shorter in magnetic regions \citep{2002ESASP.508..107V}, and the acoustic resonance between the excitation source and the upper reflection level, which is sensitive to the acoustic reflectivity of the atmosphere, changes its properties during the solar cycle in a way that is consistent with reflectivity increasing with increasing solar activity \citep{1998MNRAS.298..464V}. Indeed, the acoustic cut-off frequency, as measured using low-degree modes, has been shown to increase as solar activity increases \citep{2011ApJ...743...99J}. What our results show is how it is increasing.

Basically, the number of regions of increased acoustic cut-off frequency that are present on the Sun will depend on the number of active regions.  Therefore it can be expected that global modes of oscillation will sample a larger region of enhanced cut-off frequency near solar maximum, than at solar minimum.

We note that the ability to measure the spatial variation of the acoustic cut-off frequency has several implications:

I. Compressive waves are one of the proposed mechanisms to explain chromospheric heating \citep{2010A&A...522A..31B, 2011ApJ...735...65F, 2016ApJ...831...24K, 2017ApJ...847....5K}, and a change in acoustic reflectivity will affect the leakage of convective energy into the solar atmosphere.  Thus, a detailed characterization of $\nu_{0}$ is fundamental to estimate which frequencies can actually contribute to the heating.

II. The acoustic cut-off frequency has been shown to be a main contributor to travel-time shift measurements \citep{2010ApJ...719.1144L, 2013A&A...558A.130S, 2017A&A...604A.126F}. Therefore interpretation of the travel-time shifts will benefit from simultaneous estimations of the atmospheric cut-off frequencies, such as those carried out in this work.

III. The variation of the acoustic cut-off frequency with time has the potential to reveal locations where there are rapid changes in the properties of the atmosphere (i.e., sound speed, strength of the magnetic field, or the angle of inclination of the magnetic field). Such variations might be expected with an eruptive event such as a flare. Tentative evidence for this type of variability was shown by \citet{2004ApJ...613L.185F}. If verified, maps of acoustic cut-off frequency will provide an important data product for space weather prediction.

\section*{Acknowledgements}
The authors thank the following people for their help with the fabrication and upgrade of the MOTH II instruments, and acquiring data at the South Pole in 2016/17: Wayne Rodgers, Cynthia Giebink, William Giebink, Les Heida, Gary Nitta, and Stefano Scardigli.  This work was funded under award OPP 1341755 from the National Science Foundation. HMI data courtesy of NASA/SDO and the HMI science team. We thank the many team members who have contributed to the success of the SDO mission and particularly to the HMI instrument. And a special thanks is due to Sebastien Couvidat for providing the 2nd Fourier coefficient HMI Doppler data. We are grateful to an anonymous referee for constructive comments, which helped to improve the paper.

\bibliography{cut_off_references}

\begin{thebibliography}{}
\expandafter\ifx\csname natexlab\endcsname\relax\def\natexlab#1{#1}\fi
\providecommand{\url}[1]{\href{#1}{#1}}
\providecommand{\dodoi}[1]{doi:~\href{http://doi.org/#1}{\nolinkurl{#1}}}
\providecommand{\doeprint}[1]{\href{http://ascl.net/#1}{\nolinkurl{http://ascl.net/#1}}}
\providecommand{\doarXiv}[1]{\href{https://arxiv.org/abs/#1}{\nolinkurl{https://arxiv.org/abs/#1}}}

\bibitem[{{Audard} {et~al.}(1998){Audard}, {Kupka}, {Morel}, {Provost}, \&
  {Weiss}}]{1998A&A...335..954A}
{Audard}, N., {Kupka}, F., {Morel}, P., {Provost}, J., \& {Weiss}, W.~W. 1998,
  \aap, 335, 954.
\newblock \doarXiv{astro-ph/9712126}

\bibitem[{{Bel} \& {Leroy}(1977)}]{1977A&A....55..239B}
{Bel}, N., \& {Leroy}, B. 1977, \aap, 55, 239

\bibitem[{{Bello Gonz{\'a}lez} {et~al.}(2010){Bello Gonz{\'a}lez}, {Flores
  Soriano}, {Kneer}, {Okunev}, \& {Shchukina}}]{2010A&A...522A..31B}
{Bello Gonz{\'a}lez}, N., {Flores Soriano}, M., {Kneer}, F., {Okunev}, O., \&
  {Shchukina}, N. 2010, \aap, 522, A31, \dodoi{10.1051/0004-6361/201014052}

\bibitem[{{Chae} \& {Litvinenko}(2018)}]{2018ApJ...869...36C}
{Chae}, J., \& {Litvinenko}, Y.~E. 2018, \apj, 869, 36,
  \dodoi{10.3847/1538-4357/aaec05}

\bibitem[{{Couvidat} {et~al.}(2012){Couvidat}, {Rajaguru}, {Wachter},
  {Sankarasubramanian}, {Schou}, \& {Scherrer}}]{2012SoPh..278..217C}
{Couvidat}, S., {Rajaguru}, S.~P., {Wachter}, R., {et~al.} 2012, \solphys, 278,
  217, \dodoi{10.1007/s11207-011-9927-y}

\bibitem[{{De Moortel} \& {Browning}(2015)}]{2015RSPTA.37340269D}
{De Moortel}, I., \& {Browning}, P. 2015, Philosophical Transactions of the
  Royal Society of London Series A, 373, 20140269,
  \dodoi{10.1098/rsta.2014.0269}

\bibitem[{{Felipe} {et~al.}(2017){Felipe}, {Braun}, \&
  {Birch}}]{2017A&A...604A.126F}
{Felipe}, T., {Braun}, D.~C., \& {Birch}, A.~C. 2017, \aap, 604, A126,
  \dodoi{10.1051/0004-6361/201730798}

\bibitem[{{Felipe} {et~al.}(2011){Felipe}, {Khomenko}, \&
  {Collados}}]{2011ApJ...735...65F}
{Felipe}, T., {Khomenko}, E., \& {Collados}, M. 2011, \apj, 735, 65,
  \dodoi{10.1088/0004-637X/735/1/65}

\bibitem[{{Felipe} {et~al.}(2018){Felipe}, {Kuckein}, \&
  {Thaler}}]{2018A&A...617A..39F}
{Felipe}, T., {Kuckein}, C., \& {Thaler}, I. 2018, \aap, 617, A39,
  \dodoi{10.1051/0004-6361/201833155}

\bibitem[{{Finsterle} {et~al.}(2004){Finsterle}, {Jefferies}, {Cacciani},
  {Rapex}, \& {McIntosh}}]{2004ApJ...613L.185F}
{Finsterle}, W., {Jefferies}, S.~M., {Cacciani}, A., {Rapex}, P., \&
  {McIntosh}, S.~W. 2004, \apj, 613, L185, \dodoi{10.1086/424996}

\bibitem[{{Fleck} {et~al.}(2011){Fleck}, {Couvidat}, \&
  {Straus}}]{2011SoPh..271...27F}
{Fleck}, B., {Couvidat}, S., \& {Straus}, T. 2011, \solphys, 271, 27,
  \dodoi{10.1007/s11207-011-9783-9}

\bibitem[{{Forte} {et~al.}(2018){Forte}, {Jefferies}, {Berrilli}, {Del Moro},
  {Fleck}, {Giovannelli}, {Murphy}, {Pietropaolo}, \&
  {Rodgers}}]{2018IAUS..335..335F}
{Forte}, R., {Jefferies}, S.~M., {Berrilli}, F., {et~al.} 2018, in IAU
  Symposium, Vol. 335, Space Weather of the Heliosphere: Processes and
  Forecasts, ed. C.~{Foullon} \& O.~E. {Malandraki}, 335--339

\bibitem[{{Jefferies} {et~al.}(2006){Jefferies}, {McIntosh}, {Armstrong},
  {Bogdan}, {Cacciani}, \& {Fleck}}]{2006ApJ...648L.151J}
{Jefferies}, S.~M., {McIntosh}, S.~W., {Armstrong}, J.~D., {et~al.} 2006, \apj,
  648, L151, \dodoi{10.1086/508165}

\bibitem[{{Jefferies} {et~al.}(1997){Jefferies}, {Osaki}, {Shibahashi},
  {Harvey}, {D'Silva}, \& {Duvall}}]{1997ApJ...485L..49J}
{Jefferies}, S.~M., {Osaki}, Y., {Shibahashi}, H., {et~al.} 1997, \apj, 485,
  L49, \dodoi{10.1086/310805}

\bibitem[{{Jim{\'e}nez} {et~al.}(2011){Jim{\'e}nez}, {Garc{\'\i}a}, \&
  {Pall{\'e}}}]{2011ApJ...743...99J}
{Jim{\'e}nez}, A., {Garc{\'\i}a}, R.~A., \& {Pall{\'e}}, P.~L. 2011, \apj, 743,
  99, \dodoi{10.1088/0004-637X/743/2/99}

\bibitem[{{Kanoh} {et~al.}(2016){Kanoh}, {Shimizu}, \&
  {Imada}}]{2016ApJ...831...24K}
{Kanoh}, R., {Shimizu}, T., \& {Imada}, S. 2016, \apj, 831, 24,
  \dodoi{10.3847/0004-637X/831/1/24}

\bibitem[{{Khomenko} \& {Collados}(2009)}]{2009A&A...506L...5K}
{Khomenko}, E., \& {Collados}, M. 2009, Astronomy and Astrophysics, 506, L5,
  \dodoi{10.1051/0004-6361/200913030}

\bibitem[{{Krishna Prasad} {et~al.}(2017){Krishna Prasad}, {Jess}, {Van
  Doorsselaere}, {Verth}, {Morton}, {Fedun}, {Erd{\'e}lyi}, \&
  {Christian}}]{2017ApJ...847....5K}
{Krishna Prasad}, S., {Jess}, D.~B., {Van Doorsselaere}, T., {et~al.} 2017,
  \apj, 847, 5, \dodoi{10.3847/1538-4357/aa86b5}

\bibitem[{{Lamb}(1909)}]{1909PLMS...7..122L}
{Lamb}, H. 1909, Proc. London Math. Soc., 7, 122

\bibitem[{{Lindsey} {et~al.}(2010){Lindsey}, {Cally}, \&
  {Rempel}}]{2010ApJ...719.1144L}
{Lindsey}, C., {Cally}, P.~S., \& {Rempel}, M. 2010, \apj, 719, 1144,
  \dodoi{10.1088/0004-637X/719/2/1144}

\bibitem[{{Moretti} {et~al.}(2007){Moretti}, {Jefferies}, {Armstrong}, \&
  {McIntosh}}]{2007A&A...471..961M}
{Moretti}, P.~F., {Jefferies}, S.~M., {Armstrong}, J.~D., \& {McIntosh}, S.~W.
  2007, Astronomy and Astrophysics, 471, 961,
  \dodoi{10.1051/0004-6361:20077247}

\bibitem[{{Murawski} {et~al.}(2016){Murawski}, {Musielak}, {Konkol}, \&
  {Wi{\'s}niewska}}]{2016ApJ...827...37M}
{Murawski}, K., {Musielak}, Z.~E., {Konkol}, P., \& {Wi{\'s}niewska}, A. 2016,
  \apj, 827, 37, \dodoi{10.3847/0004-637X/827/1/37}

\bibitem[{{Nagashima} {et~al.}(2014){Nagashima}, {L{\"o}ptien}, {Gizon},
  {Birch}, {Cameron}, {Couvidat}, {Danilovic}, {Fleck}, \&
  {Stein}}]{2014SoPh..289.3457N}
{Nagashima}, K., {L{\"o}ptien}, B., {Gizon}, L., {et~al.} 2014, \solphys, 289,
  3457, \dodoi{10.1007/s11207-014-0543-5}

\bibitem[{{Rajaguru} {et~al.}(2019){Rajaguru}, {Sangeetha}, \&
  {Tripathi}}]{2019ApJ...871..155R}
{Rajaguru}, S.~P., {Sangeetha}, C.~R., \& {Tripathi}, D. 2019, \apj, 871, 155,
  \dodoi{10.3847/1538-4357/aaf883}

\bibitem[{{Rijs} {et~al.}(2015){Rijs}, {Moradi}, {Przybylski}, \&
  {Cally}}]{2015ApJ...801...27R}
{Rijs}, C., {Moradi}, H., {Przybylski}, D., \& {Cally}, P.~S. 2015, \apj, 801,
  27, \dodoi{10.1088/0004-637X/801/1/27}

\bibitem[{{Scherrer} {et~al.}(2012){Scherrer}, {Schou}, {Bush}, {Kosovichev},
  {Bogart}, {Hoeksema}, {Liu}, {Duvall}, {Zhao}, {Title}, {Schrijver},
  {Tarbell}, \& {Tomczyk}}]{2012SoPh..275..207S}
{Scherrer}, P.~H., {Schou}, J., {Bush}, R.~I., {et~al.} 2012, \solphys, 275,
  207, \dodoi{10.1007/s11207-011-9834-2}

\bibitem[{{Schou} {et~al.}(2012){Schou}, {Scherrer}, {Bush}, {Wachter},
  {Couvidat}, {Rabello-Soares}, {Bogart}, {Hoeksema}, {Liu}, {Duvall}, {Akin},
  {Allard}, {Miles}, {Rairden}, {Shine}, {Tarbell}, {Title}, {Wolfson},
  {Elmore}, {Norton}, \& {Tomczyk}}]{2012SoPh..275..229S}
{Schou}, J., {Scherrer}, P.~H., {Bush}, R.~I., {et~al.} 2012, \solphys, 275,
  229, \dodoi{10.1007/s11207-011-9842-2}

\bibitem[{{Schunker} \& {Cally}(2006)}]{2006MNRAS.372..551S}
{Schunker}, H., \& {Cally}, P.~S. 2006, \mnras, 372, 551,
  \dodoi{10.1111/j.1365-2966.2006.10855.x}

\bibitem[{{Schunker} {et~al.}(2013){Schunker}, {Gizon}, {Cameron}, \&
  {Birch}}]{2013A&A...558A.130S}
{Schunker}, H., {Gizon}, L., {Cameron}, R.~H., \& {Birch}, A.~C. 2013, \aap,
  558, A130, \dodoi{10.1051/0004-6361/201321485}

\bibitem[{{Souffrin}(1972)}]{1972A&A....17..458S}
{Souffrin}, P. 1972, \aap, 17, 458

\bibitem[{{Vorontsov}(2002)}]{2002ESASP.508..107V}
{Vorontsov}, S.~V. 2002, in ESA Special Publication, Vol. 508, From Solar Min
  to Max: Half a Solar Cycle with SOHO, ed. A.~{Wilson}, 107--110

\bibitem[{{Vorontsov} {et~al.}(1998){Vorontsov}, {Jefferies}, {Duval}, \&
  {Harvey}}]{1998MNRAS.298..464V}
{Vorontsov}, S.~V., {Jefferies}, S.~M., {Duval}, T.~L., J., \& {Harvey}, J.~W.
  1998, \mnras, 298, 464, \dodoi{10.1046/j.1365-8711.1998.01630.x}

\bibitem[{{Wi{\'s}niewska} {et~al.}(2016){Wi{\'s}niewska}, {Musielak},
  {Staiger}, \& {Roth}}]{2016ApJ...819L..23W}
{Wi{\'s}niewska}, A., {Musielak}, Z.~E., {Staiger}, J., \& {Roth}, M. 2016,
  \apj, 819, L23, \dodoi{10.3847/2041-8205/819/2/L23}

\bibitem[{{Worrall}(1991)}]{1991MNRAS.251..427W}
{Worrall}, G. 1991, \mnras, 251, 427, \dodoi{10.1093/mnras/251.3.427}

\end{thebibliography}

\end{document}